\documentstyle[aps,twocolumn,psfig]{revtex}

\begin{document}

\title{Statistical properties of twin beams generated in
spontaneous parametric downconversion}

\author{Jan Pe\v{r}ina Jr.$ {}^{a,b} $\thanks{e-mail:
perina\underline{ }j@sloup.upol.cz}, \\
Ond\v{r}ej Haderka$ {}^{a,b} $, \\
Martin Hamar$ {}^{a,b} $, \\
$ {}^{a} $ Joint Laboratory of Optics
of Palack\'{y} University and \\
Institute of Physics of Academy of Sciences of the Czech
Republic, \\
17. listopadu 50A, 772 00 Olomouc,
Czech Republic \\
$ {}^{b} $ Department of Optics, Faculty of Natural Sciences, Palack\'{y}
University, \\ 17. listopadu 50,
772 00 Olomouc, Czech Republic }

\maketitle

\begin{abstract}
Measurement of photon-number statistics of fields composed of
photon pairs generated in spontaneous parametric downcoversion
pumped by strong ultrashort pulses is described. Final detection
quantum efficiencies,
noises as well as possible loss of one or both photons from a pair are
taken into account.
Measured data provided by an intensified single-photon CCD camera are
analyzed along the developed model. The joint signal-idler photon-number
distribution is obtained
using the expectation maximization algorithm. Covariance of the
signal and idler photon-numbers equals 80~\%. Statistics of the
generated photon pairs are identified to be Poissonian in our case.
Distribution of the integrated intensities of the signal and
idler fields shows strong correlations between the fields.
Negative values of this distribution occurring
in some regions clearly demonstrate a nonclassical character of
the light composed of photon pairs.
\end{abstract}

\section{Introduction}

Light generated in the process of spontaneous parametric
downconversion is emitted in photon pairs \cite{mandelwolf}.
Photons comprising one photon pair are strongly correlated --- they
are entangled. Entanglement of photons in a pair has been used
in many experiments that have provided a deep insight into the laws
of quantum mechanics \cite{milburnwalls,perina}.
Among others, the measured violation of
Bell inequalities ruled out neoclassical local hidden-variables
theories. Photon pairs
have also found ways to practical applications, e.g., in quantum
cryptography, measurement of ultrashort time delays, or absolute
measurements of detection quantum efficiencies.
These experiments utilized photon fields that
contain only one photon pair in a measured time window with high
probability.

There have been
experiments (teleportation, measurement of GHZ correlations,
etc.) measuring triple and quadruple coincidence counts
caused by fields containing two photon pairs in the time window
given by an ultrashort pump pulse during the
last couple of years. However, states used for such experiments
contain a very low fraction of two photon-pair states in comparison with
one photon-pair state and the vacuum state. The reason is to eliminate the
influence of three and more-than-three photon-pair states in considered
experimental setups. Measurements done in such setups are
conditional and they require long data-acquisition times.

The use of more powerful pulses pump lasers as well as development
of materials with higher values of $ \chi^{(2)} $ susceptibilities
open the way to generate fields containing more than one photon
pair from one pump pulse with nonnegligible probabilities.
For such fields, the influence of states containing two photon-pairs
has to be judged even in experiments utilizing primarily just
one photon pair. For example, the effect of two-photon-pair
states in measurement of time-bin entanglement has been
analyzed in \cite{gisin}.
Contribution from more-than-one photon-pair states becomes essential
for higher mean
values of photon pairs and such fields then have to be characterized
in general by their photon-pair statistics.

Determination of photon-pair statistics is necessary also for
characterization of weak continuous downconverted fields if they
are detected in long time windows \cite{teich}. In this case
photon-pair statistics were determined to be Poissonian if
dead-time effects in their detection were eliminated \cite{teich}.

We describe the behaviour of photon pairs in a general photon-number
resolving system. Assuming detection of
generated photon pairs by an intensified CCD camera, we can
obtain a joint photon-number distribution of the signal and idler
fields at the output plane of a crystal from measured data using
a developed model of detection and a suitable reconstruction algorithm.
The generated field can then be characterized in general by
the joint distribution of the signal and
idler integrated intensities.

The paper is organized as follows. Sec. 2 contains a general
model describing a photon-number-resolving detection device. Sec. 3
is devoted to reconstruction of the joint signal-idler photon-number
distribution. Joint signal-idler integrated-intensity
distribution is determined in Sec. 4.
Sec. 5 provides conclusions. Statistical properties of fields
generated in spontaneous parametric downconversion are derived
in Appendix A from first principles.

\section{Model describing the measurement of
multiphoton-coincidence counts}

Measurement of the joint signal-idler photon-number distribution
can be in general described using the scheme
shown in Fig.~1 \cite{crypto,greece,nase}.
\begin{figure}         
 \centerline{\psfig{file=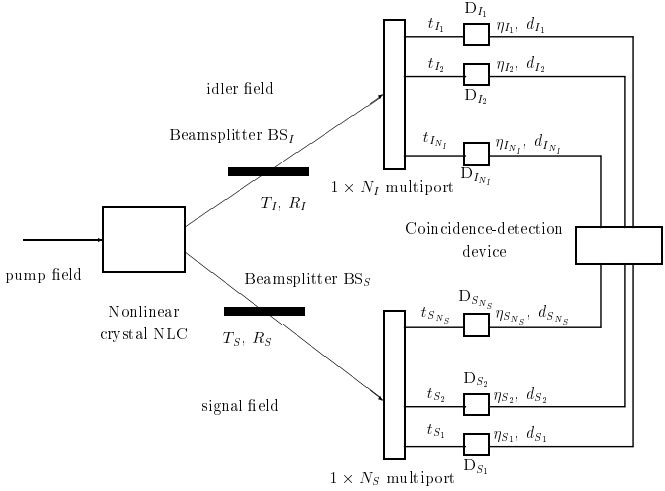,width=0.49\textwidth}}
 \vspace{3mm}
  \caption{Scheme of the considered model.
  Photon pairs are generated in nonlinear crystal
  NLC. Virtual beamsplitters $ {\rm BS}_S $ and $ {\rm BS}_I $
  describe possible losses of one or both photons from a pair before they are
  detected. Signal (idler) photons propagate through
  a $ 1\times N_S $ ($ 1\times N_I $) multiport and are detected
  in single-photon detectors $ {\rm D}_{S_1} $, $ {\rm D}_{S_2} $, \ldots,
  $ {\rm D}_{S_{N_S}} $ ($ {\rm D}_{I_1} $, $ {\rm D}_{I_2} $, \ldots,
  $ {\rm D}_{I_{N_I}} $). Signals from the detectors are registered
  in a coincidence-detection device.}
\end{figure}
Photon pairs occurring in the output plane of
nonlinear crystal NLC propagate towards photon-number-resolving
detection devices. One or both photons from a pair may be lost
before they reach
their detection devices. The reasons may be geometric filtering
(one photon from a photon pair is not collimated to the detector area),
reflections on optical elements in experimental setup or
absorption of a photon along its path to a detection device.
We describe this effect by two
beamsplitters BS$ {}_S $ and BS$ {}_I $ \cite{campos} placed
in the signal-field and idler-field paths, respectively.
We model a photon-number-resolving detection device as
a multiport $ 1\times N $ \cite{jex} followed by $ N $
single-photon detectors.
This description is applicable also when an intensified
CCD camera is used (as in our experiment). We note that detectors
able to resolve directly photon numbers to some extent have been
constructed \cite{kim,miller,brattke}. From practical viewpoint,
detectors using time multiplexing (reached in fiber optics) and
one or two single-photon detectors are promising
\cite{rehacek,achilles,fitch}.
We note that an intensified CCD camera has already occurred
to be useful when studying spatial correlations of photon pairs
\cite{jost}.

We assume that the signal and idler fields in the output plane
of nonlinear crystal NLC are described by the following
statistical operator $ \hat{\rho}_{SI} $ written in Fock basis:
\begin{equation}           
 \hat{\rho}_{SI} = \sum_{n_S=0}^{\infty} \sum_{n_I=0}^{\infty}
  p(n_S,n_I) |n_S\rangle_S {}_S \langle n_S| \otimes
  |n_I\rangle_I {}_I \langle n_I| ;
\end{equation}
the symbol $ p(n_S,n_I) $ denotes the joint
signal-idler photon-number distribution.

Statistical operator $ \hat{\rho}^D_{SI} $ appropriate for the signal
and idler fields in front of the detection devices can be written
as:
\begin{eqnarray}           
 \hat{\rho}^D_{SI} &=& \sum_{n_S=0}^{\infty} \sum_{n_I=0}^{\infty}
  p(n_S,n_I) \nonumber \\
  & & \mbox{} \times \sum_{l_S=0}^{n_S}
  \pmatrix{n_S \cr l_S\cr}  T_S^{l_S} R_S^{n_S-l_S}
|l_S\rangle_S {}_S \langle l_S| \nonumber \\
 & & \mbox{} \times \sum_{l_I=0}^{n_I}
  \pmatrix{n_I \cr l_I\cr}  T_I^{l_I} R_I^{n_I-l_I}
  |l_I\rangle_I {}_I \langle l_I|  .
  \label{2}
\end{eqnarray}
The symbols $ R_S $ and $ R_I $ ($ T_S $ and $ T_I $) denote
intensity reflectivities (transmissivities) of the beamsplitters
in the signal-field and idler-field paths.

We assume a multiport $ 1\times N_S $ ($ 1\times N_I $) followed by
$ N_S $ ($ N_I $) single-photon detectors with quantum efficiencies
$ \eta_{S_i} $
($ \eta_{I_i} $) and dark-count rates $ d_{S_i} $ ($ d_{I_i} $)
in the signal (idler) path. Detection of a photon in the $ k $-th
detector is described by the detection operator
$ \hat{D}_k $ \cite{crypto}:
\begin{equation}       
 \hat{D}_k = \sum_{n=0}^{\infty} \{ [1- (1-\eta_k)^n]
 + d_k (1-\eta_k)^n \} |n\rangle_k {}_k \langle n|  .
\end{equation}
Detection operator $ \hat{D}_k^{\rm no} $ corresponds to the case
when no detection has occurred:
\begin{equation}      
 \hat{D}_k^{\rm no} = 1 - \hat{D}_k .
\end{equation}
The effect of splitting photons in the signal field in
a $ 1\times N_S $ multiport
can be described by the relation $ \hat{a}_S = \sum_{i=1}^{N_S}
t_{S_i} \hat{a}_{S_i} $, where $ \hat{a}_S $ is the annihilation
operator of the signal field entering the multiport, the
annihilation operator $ \hat{a}_{S_i} $ describes a field at
the $ i $-th multiport output, and $ t_{S_i} $ stands for
amplitude transmissivity of a photon from the input to the
$ i $-th output. $ 1\times N_I $ multiport in the idler-field
path is described similarly and the symbol $ t_{I_i} $ then
refers to amplitude transmissivity of a photon from the input to
the $ i $-th multiport output.

The probability $ C_{{S^D},{I^D}} $ that given $ c_S $ detectors
in the signal field and given $ c_I $ detectors in the idler
field detect
a photon whereas the rest of detectors does not register a photon
is determined as follows:
\begin{eqnarray}     
C_{{S^D},{I^D}} &=& {\rm Tr}_{SI} \left\{ \hat{\rho}^D_{SI}
\prod_{a\in {S^D}} \hat{D}_a \; \prod_{b\in S \backslash {S^D}}
\hat{D}_b^{\rm no} \right. \nonumber \\
& &  \left. \times
\prod_{c\in {I^D}} \hat{D}_c \; \prod_{d\in I \backslash {I^D}}
\hat{D}_d^{\rm no} \right\}.
 \label{5}
\end{eqnarray}
The symbol $ S $ ($ I $) denotes the set of all signal-field
(idler-field) detectors $ S = \{S_1,\ldots,S_{N_S} \} $
($ I = \{I_1,\ldots,I_{N_I} \} $). The set $ S^D $ ($ I^D $)
contains signal-field (idler-field) detectors that registered
a photon.

Using the statistical operator $ \hat{\rho}^D_{SI} $ in
Eq.~(\ref{2})
the probability $ C_{{S^D},{I^D}} $ defined in Eq.~(\ref{5}) is
written in the form:
\begin{eqnarray}        
 C_{{S^D},{I^D}} &=& \sum_{n_S=0}^{\infty} \sum_{n_I=0}^{\infty}
  p(n_S,n_I) K_{S,{S^D}}(n_S) K_{I,{I^D}}(n_I) ,
   \nonumber \\
  K_{S,{S^D}}(n_S) &=& (-1)^{c_S} \left[ \prod_{b\in S}(1-d_b)
  \right] \nonumber \\
   & & \hspace{-1.7cm} \times \left[ T_S\left( \sum_{c\in S}
 |t_{c}|^2(1-\eta_c) \right) + R_S \right]^{n_S}  \nonumber \\
   & & \hspace{-1.7cm} \mbox{} +
   \frac{(-1)^{c_S-1}}{1!} \sum_{a\in {S^D}} \left[  \prod_{b\in S
   \backslash \{a\}} (1-d_b)  \right] \nonumber \\
   & & \hspace{-1.7cm} \times
  \left[ T_S\left( |t_a|^2 \eta_a + \sum_{c\in S\backslash \{a\}}
  |t_{c}|^2(1-\eta_c) \right) +  R_S \right]^{n_S} +
   \ldots   \nonumber \\
   & & \hspace{-1.7cm} \mbox{}
   + \left[ \prod_{b\in S\backslash {S^D}}(1-d_b)
   \right]  \nonumber \\
   & & \hspace{-1.7cm} \times
  \left[ T_S\left( \sum_{c\in {S^D}} |t_c|^2 + \sum_{c\in
   S\backslash {S^D}} |t_{c}|^2(1-\eta_c) \right) +
   R_S \right]^{n_S} , \nonumber \\
  K_{I,{I^D}}(n_I) &=& (-1)^{c_I} \left[ \prod_{b\in I}(1-d_b)
  \right] \nonumber \\
  & & \hspace{-1.7cm} \times
  \left[ T_I\left( \sum_{c\in I} |t_{c}|^2(1-\eta_c) \right) +
   R_I \right]^{n_I}  \nonumber \\
   & & \hspace{-1.5cm} \mbox{} +
   \frac{(-1)^{c_I-1}}{1!} \sum_{a\in {I^D}} \left[  \prod_{b\in I
   \backslash \{a\}} (1-d_b)  \right]  \nonumber \\
   & & \hspace{-1.7cm} \times
  \left[ T_I\left( |t_a|^2 \eta_a + \sum_{c\in I\backslash \{a\}}
  |t_{c}|^2(1-\eta_c) \right) +  R_I \right]^{n_I} +
   \ldots  \nonumber \\
   & & \hspace{-1.7cm} \mbox{}
   + \left[ \prod_{b\in I\backslash {I^D}}(1-d_b)
   \right] \nonumber \\
   & & \hspace{-1.7cm} \times
  \left[ T_I\left( \sum_{c\in {I^D}} |t_c|^2 + \sum_{c\in
   I\backslash {I^D}} |t_{c}|^2(1-\eta_c) \right) +
   R_I \right]^{n_I} . \nonumber \\
  & &
\end{eqnarray}

We now consider symmetric multiports ($ t_{S_1}= t_{S_2}= \ldots
t_{S_{N_S}} = t_S $, $ t_{I_1}=t_{I_2}= \ldots = t_{I_{N_I}} = t_I $)
and detectors having the same characteristics in the signal
and idler fields ($ \eta_{S_1} =\eta_{S_2} = \ldots = \eta_{S_{N_S}}
= \eta_S $,
$ d_{S_1} = d_{S_2} = \ldots = d_{S_{N_S}}= d_S $,
$ \eta_{I_1} = \eta_{I_2} = \ldots = \eta_{I_{N_I}} = \eta_I $,
$ d_{I_1} = d_{I_2} = \ldots = d_{I_{N_I}}= d_I $).
Then the probability $ f^{N_S,N_I}(c_S,c_I) $ of having $ c_S $
detections somewhere at $ N_S $ signal detectors and
$ c_I $ detections somewhere at $ N_I $ idler detectors can be
expressed as:
\begin{equation}      
 f^{N_S,N_I}(c_S,c_I) = \pmatrix{N_S \cr c_S \cr}
 \pmatrix{N_I \cr c_I \cr} C_{{S^D},{I^D}} .
\end{equation}
Using the expression for $ C_{{S^D},{I^D}} $ in Eq. (6)
we arrive at the relation:
\begin{eqnarray}      
 f^{N_S,N_I}(c_S,c_I) &=& \sum_{n_S=0}^{\infty} \sum_{n_I=0}^{\infty}
 p(n_S,n_I) \nonumber \\
 & & \mbox{} \times  K^{S,N_S}(c_S,n_S) K^{I,N_I}(c_I,n_I) ,
 \label{8}
\end{eqnarray}
where
\begin{eqnarray}     
  K^{i,N_i}(c_i,n_i) &=& \pmatrix{N_i \cr c_i\cr} (1-d_i)^{N_i}
  (1-T_i \eta_i)^{n_i} (-1)^{n_i} \nonumber \\
   & & \hspace{-1.5cm} \times \sum_{l=0}^{c_i}
  \pmatrix{c_i \cr l \cr} \frac{(-1)^l}{(1-d_i)^l}
  \left( 1 + \frac{l}{N_i} \frac{T_i\eta_i}{1-T_i\eta_i}
   \right)^{n_i} , \nonumber \\
   & & \hspace{1cm} i=S,I .
 \label{9}
\end{eqnarray}

If the number of photons detected by an intensified CCD camera
is much lower than the number of pixels detecting the field
with a nonnegligible probability,
the limits $ N_S \longrightarrow \infty $ and
$ N_I \longrightarrow \infty $ are appropriate.
When determining these limits, the overall noise levels
$ D_S $ and $ D_I $ are hold to be constant ($ D_S = N_S d_S $,
$ D_I = N_I d_I $). The coefficients $ K $ defined in
Eq.~(\ref{9}) then simplify:
\begin{eqnarray}     
  K^{i,\infty}(c_i,n_i) &=& \sum_{l=0}^{\min (c_i,n_i)}
  \pmatrix{n_i \cr l\cr} (T_i\eta_i)^l (1-T_i\eta_i)^{n_i-l}
  \nonumber \\
  & & \mbox{} \times
   \frac{ D_i^{c_i-l}}{(c_i-l)!}  \exp(-D_i) , \hspace{0.8cm} i=S,I .
  \label{10}
\end{eqnarray}

\section{Reconstruction of the joint signal-idler photon-number
distribution}

The probabilities (frequencies)
$ f^{\infty,\infty}(c_S,c_I) $ are measured and
the relation in Eq.~(\ref{8}) then has to be inverted in order to
obtain the joint signal-idler photon-number distribution $ p(n_S,n_I) $.
The relation in Eq. (\ref{8}) together with the coefficients
$ K^{i,\infty}(c_i,n_i) $ defined in Eq. (\ref{10}) can be
inverted under special conditions analytically.
For instance, if $ D_S = D_I = 0 $ the inversion relation
is found using ``convolution''of $ f^{\infty,\infty} $
with the Bernoulli distributions
with efficiencies $ 1/(T_S\eta_S) $ and $ 1/(T_I\eta_I) $.
However, analytical approaches are not
suitable for processing real experimental data
\cite{mogilevtsev}.
Reconstruction algorithms have occurred to be suitable in this
case \cite{rehacek}.  Such algorithms are able to find a joint
signal-idler photon-number distribution $ \rho^{(\infty)}(n_S,n_I) $ that
gives the measured frequencies
$ f^{\infty,\infty}(c_S,c_I) $ with the highest probability
from all possible photon-number distributions. We use
the Kullback-Leibler divergence as a measure of distance between
the experimental data and results provided by the developed theory. The joint
signal-idler photon-number distribution $ \rho^{(\infty)}(n_S,n_I) $
that minimizes the Kullback-Leibler divergence may be found,
e.g., using the iterative Expectation-Maximization algorithm
\cite{dempster,vardi}:
\begin{eqnarray}      
\rho^{(n+1)}(n_S,n_I) &=& \rho^{(n)}(n_S,n_I)
  \nonumber \\
 & & \hspace{-2.8cm} \times \sum_{i_S,i_I=0}^{\infty}
  \frac{ f^{\infty,\infty}(i_S,i_I)
K^{S,\infty}(i_S,n_S) K^{I,\infty}(i_I,n_I) }{
\sum_{j_S,j_I=0}^{\infty} K^{S,\infty}(i_S,j_S)
K^{I,\infty}(i_I,j_I) \rho^{(n)}(j_S,j_I) } .
 \nonumber \\
 & &
\label{11}
\end{eqnarray}
The symbol $ \rho^{(n)}(n_S,n_I) $ denotes the joint
signal-idler photon-number distribution after
the $ n $-th step of iteration, $ \rho^{(0)}(n_S,n_I) $ is an
arbitrary initial photon-number distribution.

Experimental setup used for the measurement of frequencies
$ f^{\infty,\infty}(c_S,c_I) $ is shown in Fig.~2.
\begin{figure}          
 \centerline{\psfig{file=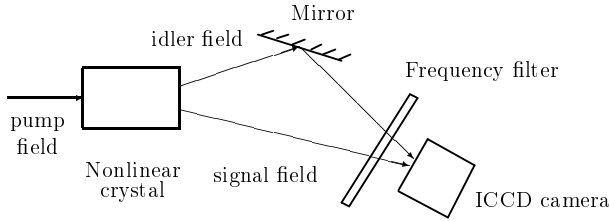,width=0.45\textwidth}}
 \vspace{3mm}
  \caption{Scheme of the setup for detection of photon pairs.
  Fields composed of typically tens of photon pairs are generated
  in a nonlinear crystal, then they propagate through a frequency
  filter and are detected on an intensified single-photon CCD
  camera.}
\end{figure}
Photon pairs are generated in a 5 mm long LiIO$ {}_3 $ crystal
pumped by intense ultrashort pulses
delivered by a titanium-sapphire femtosecond laser followed
by a regenerative amplifier at the wavelength of 800~nm.
The laser system runs at the repetition rate of 11~kHz and
after converting the 800-nm beam to its second harmonic
it typically delivers pulses with the energy of 0.1 $\mu$J.
Photons in the idler field are reflected from a high-reflectivity
mirror and impinge on an intensified CCD camera. Signal photons
come directly to the photocathode
of the intensified camera. Three regions of interest are defined
in the field of view of the camera: the first one is for the signal
field, the second one for the idler field, and the third region
serves for reference measurements of the noise level.
The whole field of the camera is filtered by a high-transmittance
high-pass filter blocking light below 700~nm
and by an intereference filter of 10~nm FWHM centered at 800~nm.
The interference filter selects nearly degenerate photon pairs.
A typical histogram of the measured frequencies of
the joint signal-idler photon-number
distribution $ f^{\infty,\infty}(c_S,c_I) $
is shown in Fig.~3.
\begin{figure}      
 \vbox{\noindent (a)
 \centerline{\psfig{file=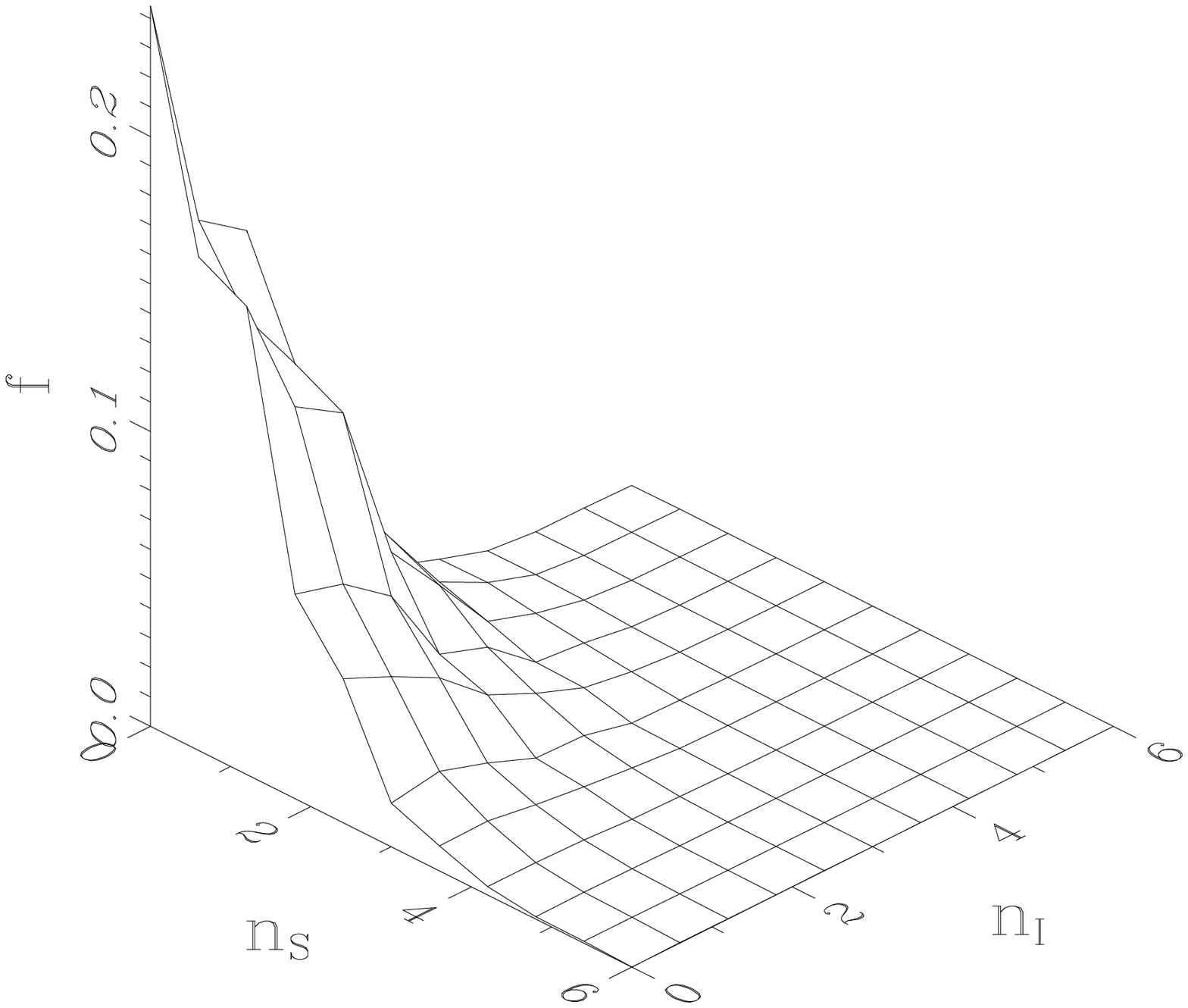,width=0.4\textwidth}}
 (b)
 \centerline{\psfig{file=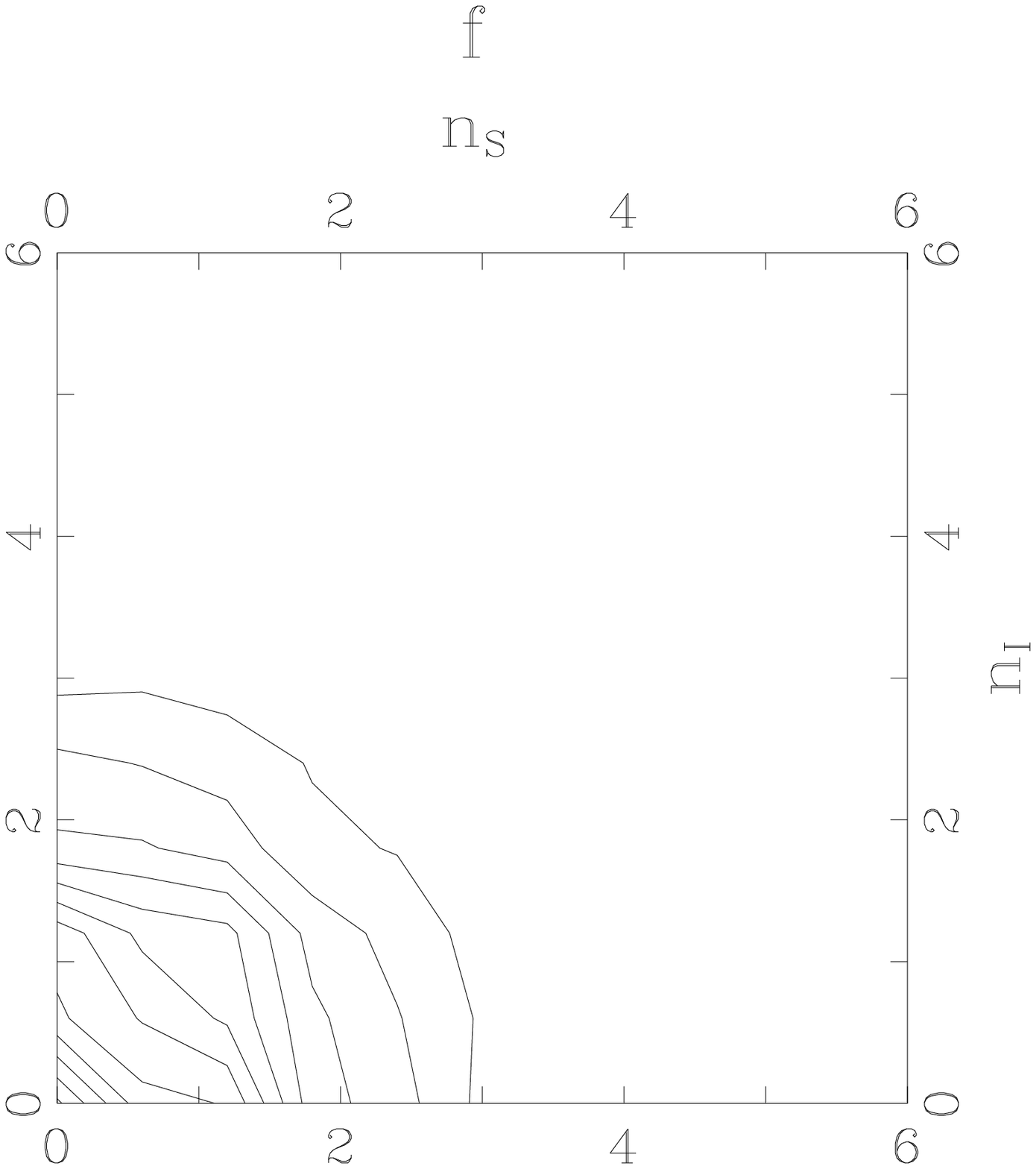,width=0.25\textwidth}}
  }
 \vspace{3mm}
  \caption{(a) Measured histogram of the frequencies
  $ f^{\infty,\infty} $ as a function of the signal
  ($ n_S $) and idler ($ n_I $) photon numbers.
  (b) Topological graph of the histogram
  $ f^{\infty,\infty} $.}
\end{figure}

The reconstruction algorithm in Eq. (\ref{11}) lead to
the joint signal-idler photon-number
distribution $ \rho^{(\infty)}(n_S,n_I) $ in the output plane of
the crystal as shown in Fig. 4. We have assumed
$ T_S \eta_S = T_I \eta_I = 0.03 $ after taking into account
losses in the setup (frequency filters, reflection on the output
plane of the crystal, final quantum efficiency of the intensified CCD camera).
The measurement has also provided values of noises;
$ D_S = D_I = 0.1 $. The initial joint signal-idler photon-number
distribution $ \rho^{(0)} $ was assumed to be uniform.
\begin{figure}      
 \vbox{\noindent (a)
 \centerline{\psfig{file=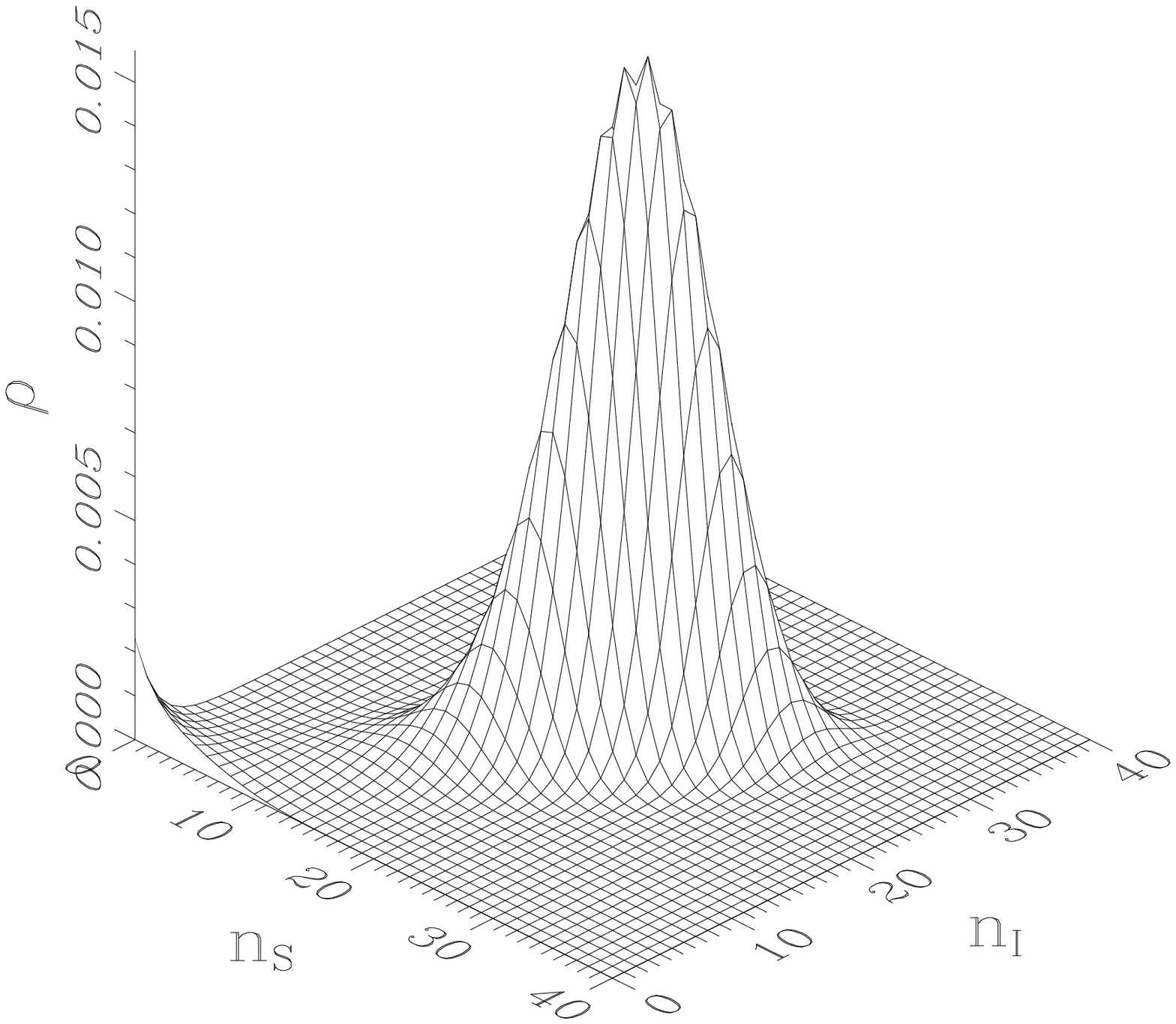,width=0.4\textwidth}}
 (b)
 \centerline{\psfig{file=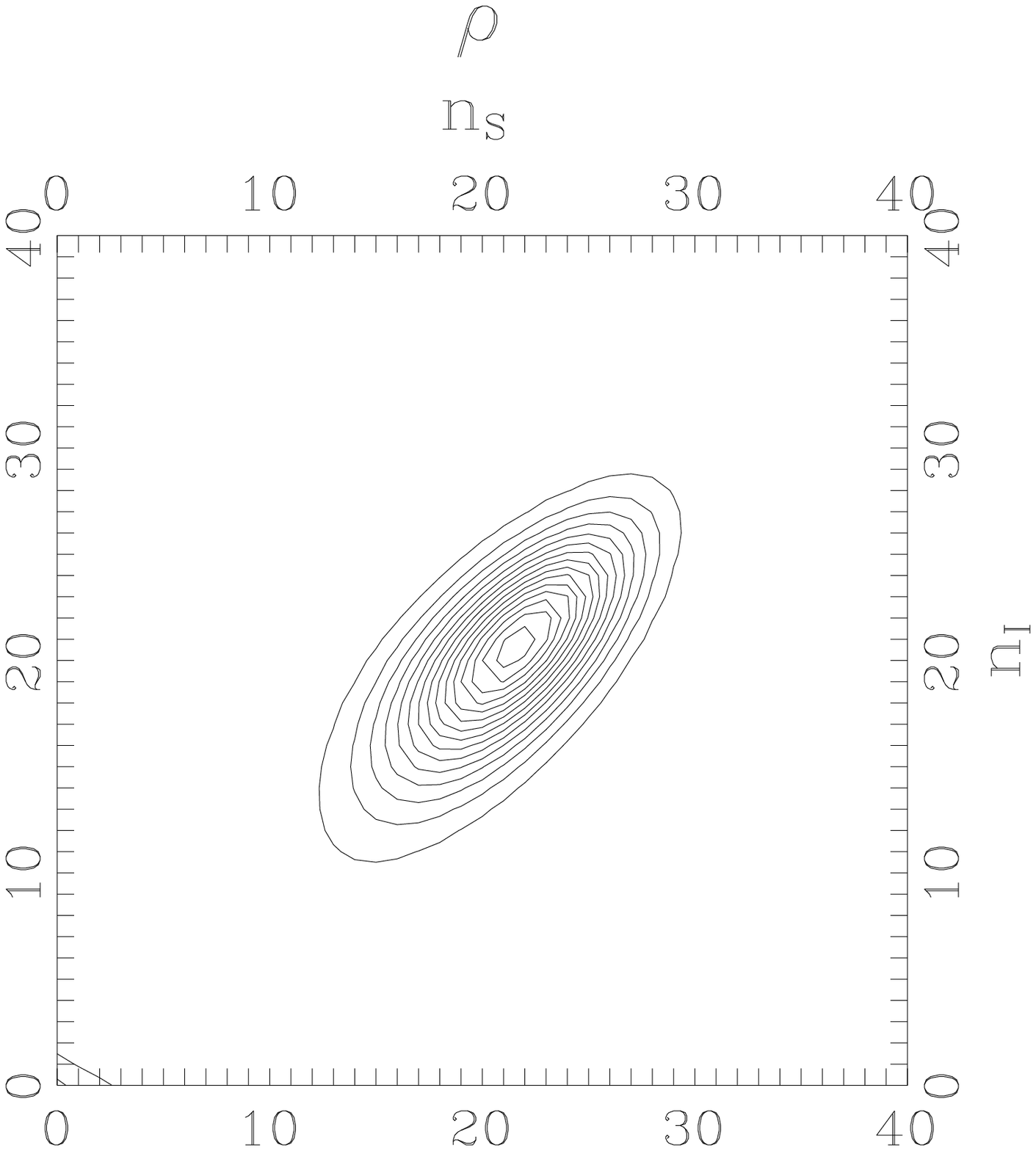,width=0.25\textwidth}}
   }
 \vspace{3mm}
  \caption{(a) Reconstructed joint signal-idler photon-number
  distribution $ \rho^{(\infty)} $ as a function of the signal
  ($ n_S $) and idler ($ n_I $) photon numbers.
  (b) Topological graph of the distribution
  $ \rho^{(\infty)} $.}
\end{figure}

The graph in Fig.~4 clearly shows that signal and idler photons
are generated in pairs to some extent. In order to quantify this
property we use covariance $ C_p $ of two stochastic signals $ n_1 $
and $ n_2 $ described by a joint probability distribution $ p(n_1,n_2) $:
\begin{eqnarray}          
  C_p &=& \frac{ \langle \Delta n_1 \Delta n_2 \rangle }{
  \sqrt{ \langle (\Delta n_1)^2
    \rangle \langle (\Delta n_2)^2 \rangle }} ,
      \label{12} \\
    & & \Delta n_i = n_i - \langle n_i \rangle,
      \vspace{1cm} i=1,2, \nonumber\\
   \langle n_1 n_2 \rangle &=& \sum_{n_1=0}^{\infty} \sum_{n_2=0}^{\infty}
     n_1 n_2 p(n_1,n_2) , \nonumber \\
   \langle n_i^m \rangle &=& \sum_{n_1=0}^{\infty} \sum_{n_2=0}^{\infty}
     n_i^m p(n_1,n_2) , \vspace{1cm} i=1,2, \;  m=1,2  \nonumber .
\end{eqnarray}
Covariance $ C_f $ of the measured distribution $ f^{\infty,\infty} $
equals 0.025 in our experiment. This value of $ C_f $ is in agreement
with the
theoretical value determined along the model developed in Sec. 2 for
values of parameters defined above. The obtained low value of $ C_f $
shows that photons from photon pairs are lost with a high probability
owing to
low detection efficiencies. However, covariance $ C_\rho $ of
the reconstructed joint signal-idler photon-number distribution
$ \rho^{(\infty)} $ equals 0.81. This means, that the reconstruction
algorithm is able to reveal clearly the original pairing of
photons.
We note that we have checked that the reconstruction
algorithm cannot give correlations of signals being
independent (i.e. with no correlation) in the input to this algorithm.
Independent simulation data provided by the
theoretical model in Sec. 2 were used for this purpose.
There are two reasons why the value of $ C_{\rho} $ characterizing
the reconstructed photon-number distribution $ \rho^{(\infty)} $
does not approach value 1. Impossibility of the developed model to
describe precisely all noises occurring in the experiment represents
the first reason. The second reason lies in the fact that
the numerical implementation of the reconstruction algorithm
looses its precision with the increasing number of iterations.

Pairing of the signal and idler photons results in narrowing
of the distribution $ \rho^{(\infty)}_{-} $ of the difference
$ n_S - n_I $ of the signal and idler photon numbers. Figure~5 shows
the distribution $ \rho^{(\infty)}_{-} $ determined
from the reconstructed $ \rho^{(\infty)}(n_S,n_I) $ and compares
it with the distribution $ \rho^{\rm indep}_{-} $ characterizing
independent marginal signal and idler distributions:
\begin{eqnarray}         
 \rho^{(\infty)}_{-}(n) &=& \sum_{n_S=0}^{\infty} \sum_{n_I=0}^{\infty}
   \delta_{n,n_S-n_I} \rho^{(\infty)}(n_S,n_I) , \nonumber  \\
 \rho^{\rm indep}_{-}(n) &=& \sum_{n_S=0}^{\infty} \sum_{n_I=0}^{\infty}
   \delta_{n,n_S-n_I} \rho^{(\infty)}_S(n_S)
   \rho^{(\infty)}_I(n_I) ,
\end{eqnarray}
and
\begin{eqnarray}         
 \rho^{(\infty)}_S(n_S) &=& \sum_{n_I=0}^{\infty}
   \rho^{(\infty)}(n_S,n_I) , \nonumber \\
 \rho^{(\infty)}_I(n_I) &=& \sum_{n_S=0}^{\infty}
   \rho^{(\infty)}(n_S,n_I) .
  \label{14}
\end{eqnarray}
The symbol $ \delta $ stands for Kronecker delta.
\begin{figure}      
 \centerline{\psfig{file=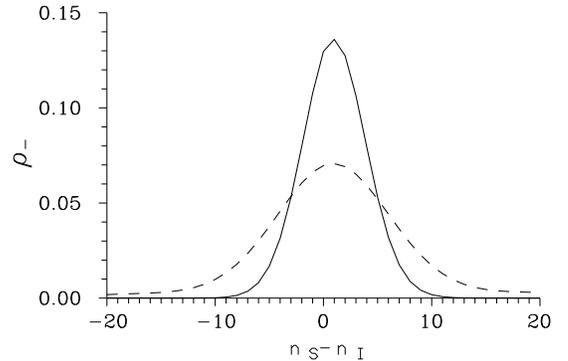,width=0.4\textwidth}}
 \vspace{3mm}
  \caption{Distributions $ \rho^{(\infty)}_{-} $ (solid curve)
   and $ \rho^{\rm indep}_{-} $ (dashed curve)
   of the difference $ n_S - n_I $
   of the signal and idler photon numbers.}
\end{figure}
We note that narrowing of the distribution $ \rho^{(\infty)}_{-} $
has been observed for stronger fields generated by parametric
downconversion in a resonator utilizing correlations of
photocurrents from two detectors \cite{zhang}.

Narrowing of the distribution $ \rho^{(\infty)}_{-} $
is accompanied by broadening
of the distribution $ \rho^{(\infty)}_{+} $ for the sum
$ n_S + n_I $:
\begin{equation}         
 \rho^{(\infty)}_{+}(n) = \sum_{n_S=0}^{\infty} \sum_{n_I=0}^{\infty}
   \delta_{n,n_S+n_I} \rho^{(\infty)}(n_S,n_I) .
\end{equation}
This stemms from the following expression for the
variances $ \langle [\Delta (n_S-n_I)]^2 \rangle $ and
$ \langle [\Delta (n_S+n_I)]^2 \rangle $:
\begin{eqnarray}         
 \langle [\Delta (n_S \pm n_I)]^2 \rangle &=&
   \langle (\Delta n_S)^2 \rangle + \langle (\Delta n_I)^2 \rangle
   \nonumber \\
   & & \mbox{} \pm 2 \sqrt{ \langle (\Delta n_S)^2 \rangle
  \langle (\Delta n_I)^2 \rangle } C_\rho ;
\end{eqnarray}
covariance $ C_\rho $ is given in Eq.~(\ref{12}).
Figure~6 shows the distribution $ \rho^{(\infty)}_{+} $ derived from
the reconstructed data together with the distribution
$ \rho^{\rm indep}_{+} $ determined from the independent
distributions $ \rho^{(\infty)}_S $ and
$ \rho^{(\infty)}_I $ (\ref{14}):
\begin{equation}        
 \rho^{\rm indep}_{+}(n) = \sum_{n_S=0}^{\infty} \sum_{n_I=0}^{\infty}
   \delta_{n,n_S+n_I} \rho^{(\infty)}_S(n_S)
   \rho^{(\infty)}_I(n_I) .
\end{equation}
\begin{figure}      
 \centerline{\psfig{file=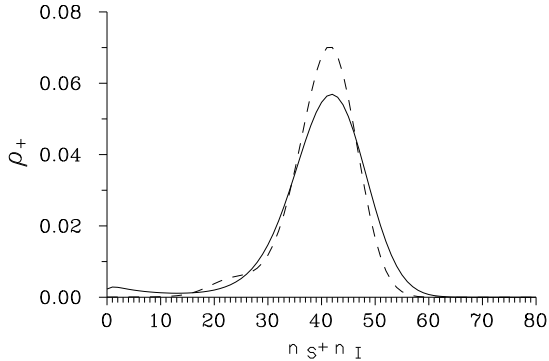,width=0.4\textwidth}}
 \vspace{3mm}
  \caption{Distributions $ \rho^{(\infty)}_{+} $ (solid curve)
   and $ \rho^{\rm indep}_{+} $ (dashed curve)
   of the sum $ n_S + n_I $
   of the signal and idler photon numbers.}
\end{figure}

Statistics of generated photon pairs as they occur in the output
plane of the crystal cannot be inferred directly from values of diagonal
elements of the reconstructed joint signal-idler photon-number distribution
$ \rho^{(\infty)} $
in Eq.~(\ref{11}) because of its blurring. This statistics can be
determined from
the marginal photon-number distributions $ \rho^{(\infty)}_S $
and $ \rho^{(\infty)}_I $ or from the photon-number
distribution of the mixed signal and idler field
$ \rho^{(\infty)}_{+} $.
Statistics are then judged according to the value of the
coefficient $ S_p $,
\begin{eqnarray}     
 S_p &=& \frac{ \langle n^2 \rangle }{ \langle n\rangle^2} -
   \frac{1}{\langle n\rangle} , \\
 \langle n^k \rangle &=& \sum_{n=0}^{\infty} n^k p(n) ,
  \vspace{1cm} k=1,2 \nonumber ;
\end{eqnarray}
$ S_p = 0 $ characterizes Poissonian statistics whereas
$ S_p = 1 $ stands for Gaussian statistics.
The values $ S_{f,S} = 1.001 $, $ S_{f,I} = 1.007 $, and
$ S_{f,+} =  1.019 $
characterize the directly measured distribution
$ f^{(\infty,\infty)} $. The reconstructed distribution
$ \rho^{(\infty)} $ is characterized by
the values $ S_{\rho,S} = 1.003 $, $ S_{\rho,I} = 1.013 $,
and $ S_{\rho,+} = 1.027 $.

Assuming Poissonian statistics of photon pairs, i.e.
$ p(n_S,n_I) = \delta_{n_S,n_I} \mu^{n_S}
\exp(-\mu)/ n_S! $, the theoretical model gives for $ \mu = 20 $
$ S_{p,S} = 1 $, $ S_{p,I} = 1  $, $ S_{p,+} = 1.025  $ in
the output plane of the crystal, whereas
$ S_{f,S} = 1 $, $ S_{f,I} = 1  $, $ S_{f,+} = 1.017 $
are appropriate for distributions detected by an intensified
CCD camera.
On the other hand, Gaussian photon-pair statistics,
$ p(n_S,n_I) = \delta_{n_S,n_I}
\mu^{n_S}/(\mu + 1)^{n_S+1} $, used in the theoretical
model with $ \mu = 20 $, give $ S_{p,S} = 2 $, $ S_{p,I} = 2  $,
$ S_{p,+} = 2 $ in the output plane of the crystal and
$ S_{f,S} = 1.69 $, $ S_{f,I} = 1.69 $, $ S_{f,+} = 1.71 $
for distributions detected by an intensified CCD camera.
Comparison of these theoretical results with the above
ones stemming from experiment
and reconstruction clearly shows that photon pairs are generated
with Poissonian statistics.

Poissonian statistics correspond to the physical picture in
which photon pairs can be generated into a great number
of independent modes distinguishable in space and time
(see Appendix A).
In case of our experimentally generated fields containing
typically several tens of photon pairs, each pair occupies its
own mode with a high probability.
If photon pairs are generated into one
well-defined space-time mode, statistics of photon pairs
would be Gaussian (see Appendix A).

\section{Joint signal-idler integrated-intensity distribution}

A complete characterisation of the generated signal and idler fields
is given by the joint distribution of their integrated intensities
$ W_S $ and $ W_I $ in the output plane of the crystal;
\begin{equation}      
 \hat{W}_l = \int_{-\infty}^{\infty} d\tau \hat{E}^{(-)}_l(\tau)
 \hat{E}^{(+)}_l(\tau) , \hspace{1cm} l=S,I .
  \label{17}
\end{equation}
The operator amplitudes $ \hat{E}^{(+)}_l $ and
$ \hat{E}^{(-)}_l $ in Eq. (\ref{17}) are given by Eq.~(\ref{A5})
in Appendix A.
Integration over time variable in Eq. (\ref{17})
is carried out over the field generated by one pump pulse.

Inverting the photodetection equation
the joint signal-idler integrated-intensity distribution
$ P(W_S,W_I,s) $ related to $ s-$ordering of field operators
can be determined using the joint signal-idler
photon-number distribution $ \rho^{(\infty)} $ as
(\cite{perinabook}, chap. 4):
\begin{eqnarray}       
 P(W_S,W_I,s) &=& \frac{4}{(1-s)^2}
 \exp\left[ - \frac{2(W_S + W_I)}{1-s} \right]
 \nonumber \\
 & & \hspace{-1cm} \times
 \sum_{n_S=0}^{\infty} \sum_{n_I=0}^{\infty}
 \frac{\rho^{(\infty)}(n_S,n_I)}{n_S! \, n_I!}
 \left( \frac{s+1}{s-1}\right)^{n_S+n_I}
  \nonumber \\
 & & \hspace{-1cm} \times
 L^0_{n_S} \left( \frac{4W_S}{1-s^2} \right)
 L^0_{n_I} \left( \frac{4W_I}{1-s^2} \right) ;
\end{eqnarray}
$ L^0_n $ are Laguerre polynomials.
The parameter $ s $ equals -1, 0, and 1 for antinormal,
symmetric, and normal ordering of field operators, respectively.
The integrated-intensity distribution related to normal ordering
does not include noise from vacuum fluctuations and that is why it shows
all features of the generated fields in their full complexity.
However, its structure is rather complex in our case of
nonclassical correlated
fields (generalized functions are necessary for the description)
and that is why we use the distribution related to symmetric
ordering. This distribution behaves well and still has the
capability to resolve nonclassical properties of the fields
by its negative values.

\begin{figure}      
\vbox{\noindent (a)
 \centerline{\psfig{file=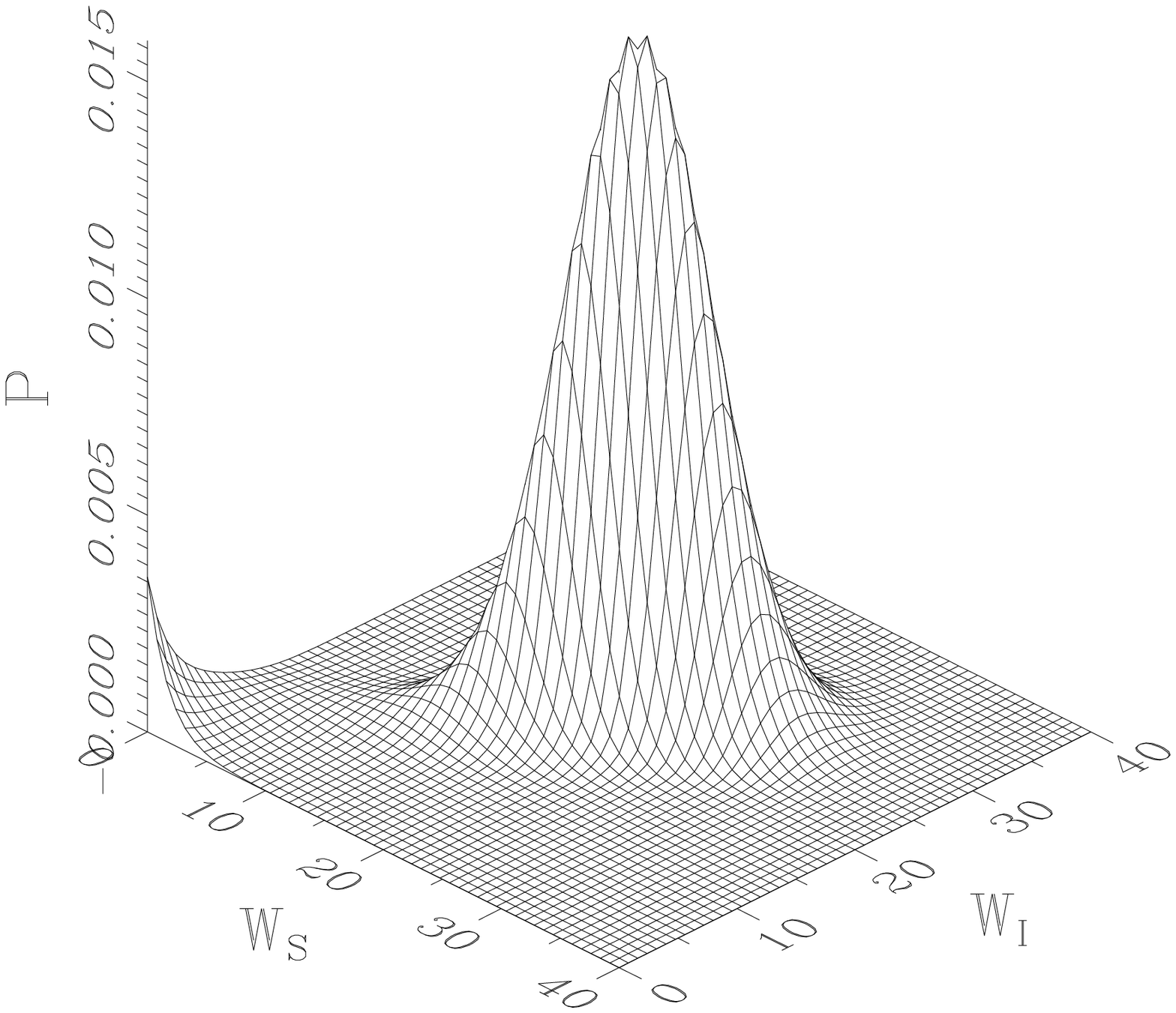,width=0.4\textwidth}}
 (b)
 \centerline{\psfig{file=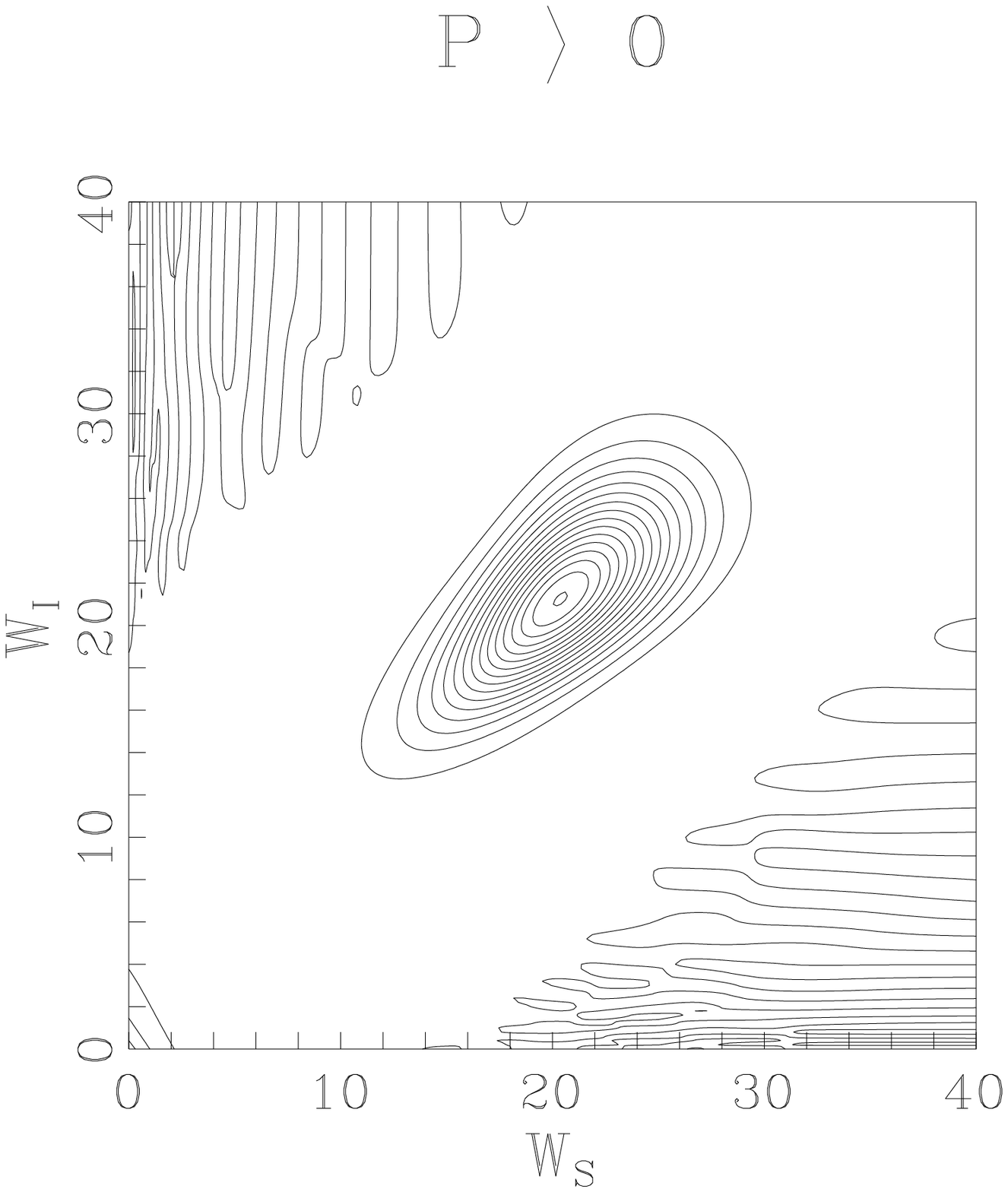,width=0.25\textwidth}}
 (c)
 \centerline{\psfig{file=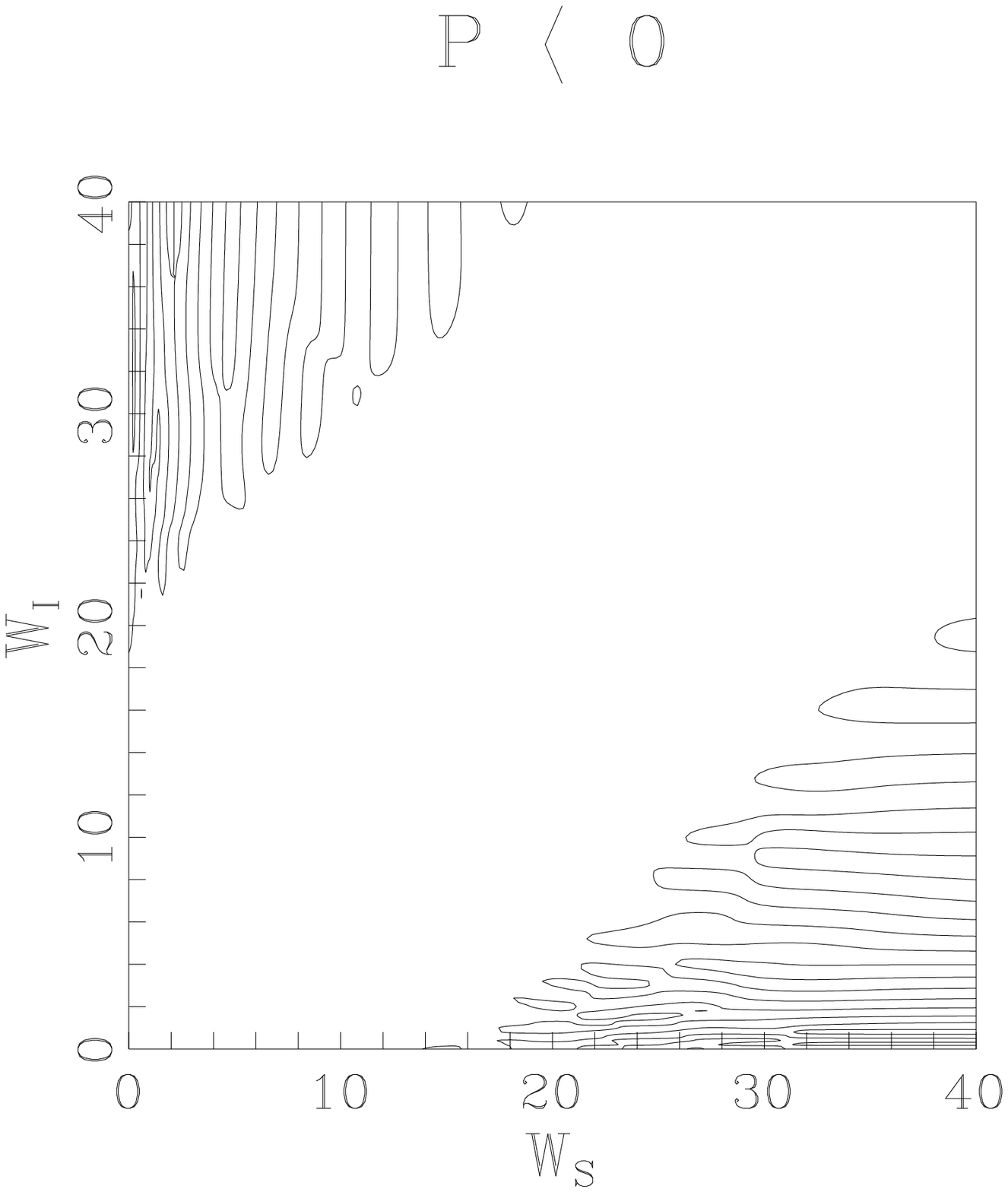,width=0.25\textwidth}}
 \vspace{3mm}
 \caption{Joint signal-idler integrated-intensity distribution
  $ P(W_S,W_I,0) $, (a) 3D plot,
  (b) topological plot showing lines having positive
  values, and (c) topological plot showing lines having negative
  values of the integrated intensity for the reconstructed
  photon-number distribution
  $ \rho^{(\infty)}(n_S,n_I) $.}
}
\end{figure}
The signal-idler integrated-intensity distribution
$ P(W_S,W_I,0) $ related
to symmetric ordering of operators and obtained from the reconstructed
signal-idler photon-number distribution $ \rho^{(\infty)} $
is shown in Fig.~7. It clearly shows strong correlations
in values of the integrated
intensities $ W_S $ and $ W_I $. This correlation
inherent to the fields in the moment of their generation again
shows that photons are generated
in pairs, i.e., the same number of energy quanta is put into both
fields during their generation. Negative values of the distribution
$ P(W_S,W_I,0) $
(see Fig. 7c) reached in some regions of values of integrated
intensities $ W_S $ and $ W_I $ mean that the generated signal-idler field
with its correlations is nonclassical (i.e., it cannot be described
by classical statistical optics).
For comparison, Fig. 8 contains the signal-idler integrated-intensity
distribution $ P(W_S,W_I,0) $ for an ideally correlated
signal-idler field with Poissonian statistics.
\begin{figure}      
\vbox{\noindent (a)
 \centerline{\psfig{file=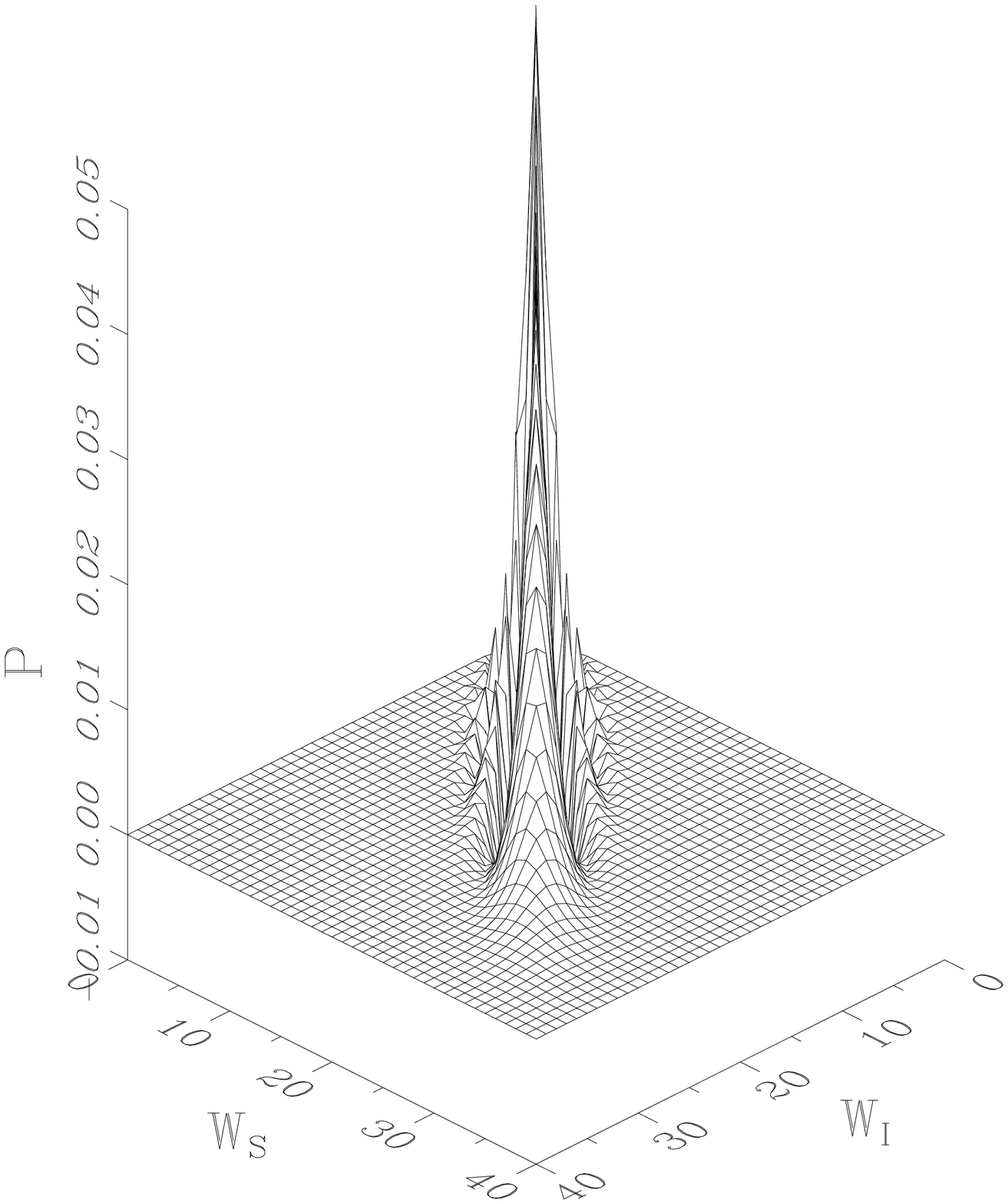,width=0.3\textwidth}}
 (b)
 \centerline{\psfig{file=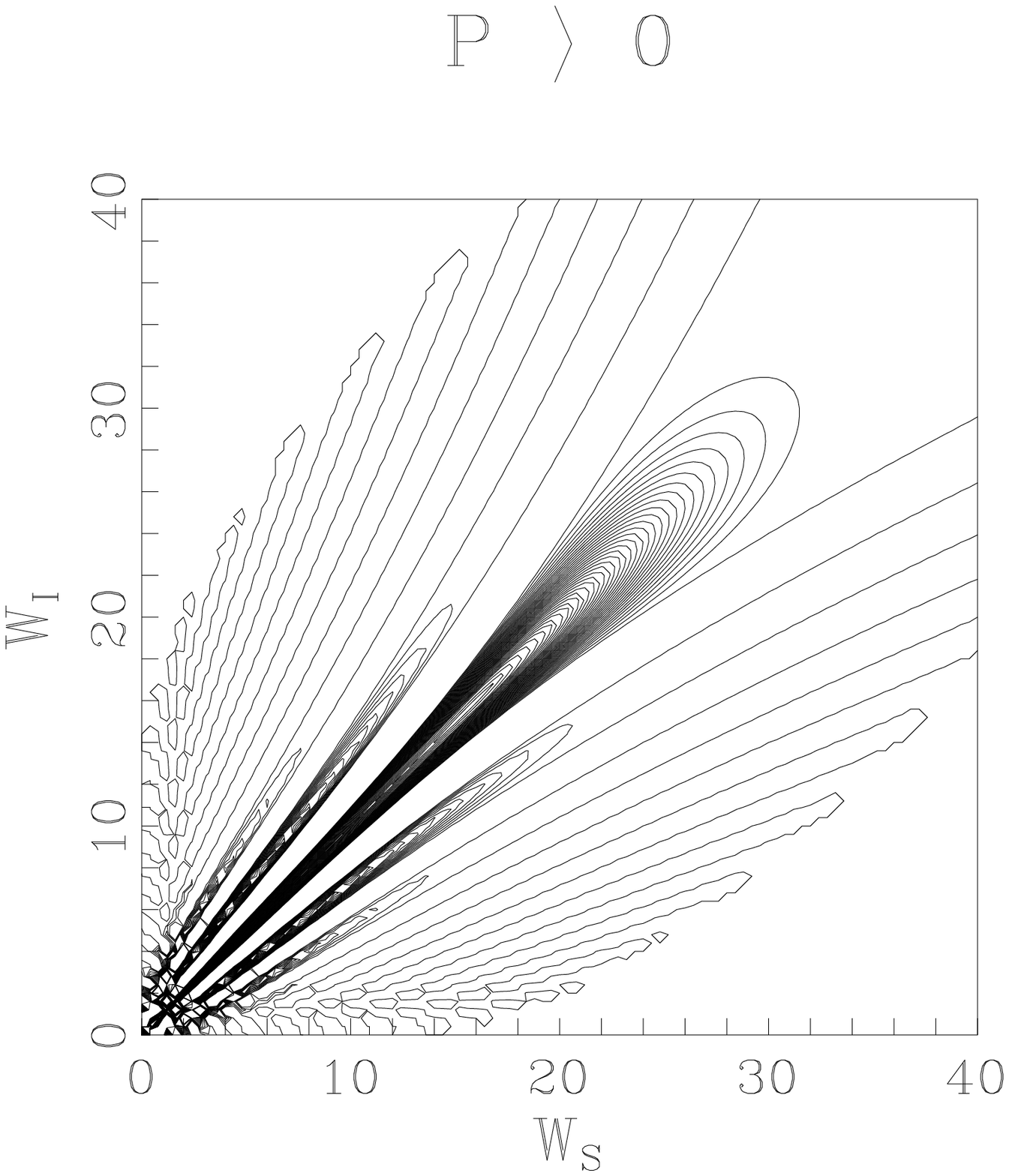,width=0.25\textwidth}}
 (c)
 \centerline{\psfig{file=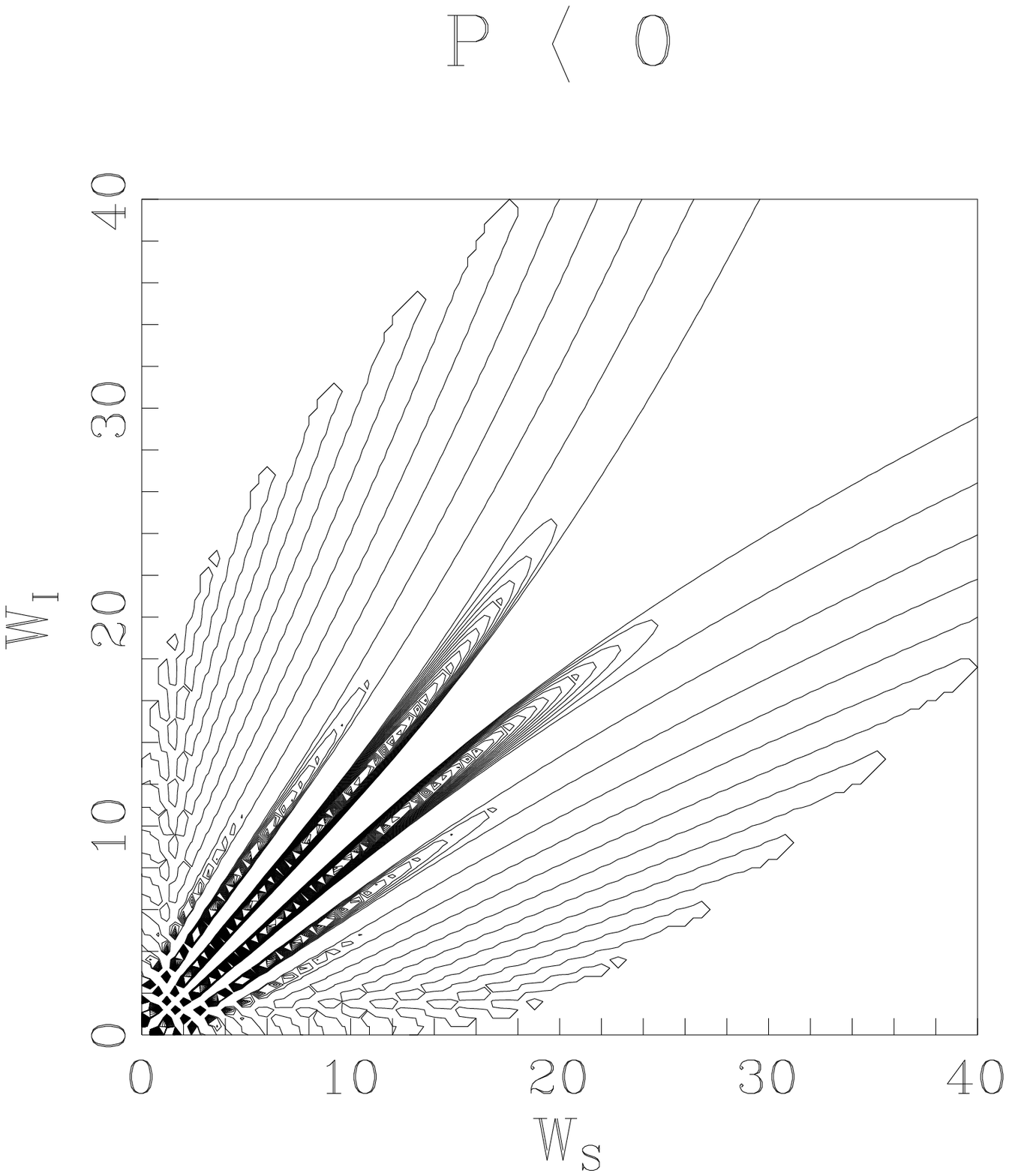,width=0.25\textwidth}}
    }
 \vspace{3mm}
  \caption{Joint signal-idler integrated-intensity distribution
  $ P(W_S,W_I,0) $, (a) 3D plot,
  (b) topological plot showing lines having positive
  values, and (c) topological plot showing lines having negative
  values for $ \rho^{(\infty)}(n_S,n_I) = \delta_{n_S,n_I}
  \mu^{n_S} \exp(-\mu) / n_S! $, $ \mu = 20 $.}
\end{figure}
All nonclassical features visible in this distribution
in Fig.~8 are
contained in the experimentally-obtained distribution shown in
Fig.~7.

\section{Conclusions}
We have experimentally verified that photons are generated
in pairs into the signal and idler fields in the nonlinear
process of spontaneous parametric downconversion.
Both the experimentally determined joint signal-idler
photon-number distribution and joint signal-idler
integrated-intensity distribution show strong correlations
between quantities characterizing the signal and idler
fields. For example, covariance of the signal and idler
photon numbers is greater than 80~\%. Statistics of the
generated photon pairs have been determined to be Poissonian.
Nonclassical character of the generated signal-idler field
manifests itself by negative values of the joint signal-idler
integrated-intensity distribution.

\section*{Acknowledgments}
The authors thank J. Pe\v{r}ina for helpful discussions.
The authors acknowledge the support by the projects Research
Center for Optics (LN00A015) and CEZJ-14/98 of
the Ministry of Education of the Czech Republic.

\appendix

\section{Photon-pair statistics in spontaneous parametric
frequency downconversion from first principles}

The interaction Hamiltonian $ \hat{H}_{\rm int} $ of the process
of spontaneous parametric frequency downconversion can be written
in the form \cite{mandelwolf,disper,perinajr}:
\begin{eqnarray}     
 \hat{H}_{\rm int}(t) &=& C_{\rm int}
 \sum_{k_s}
 \sum_{k_i} \phi(k_S,k_I,t)
 \hat{a}_S^{\dagger}(k_S)
 \hat{a}_I^{\dagger}(k_I)
 + \mbox{H.c.} \nonumber \\
 &=& \hat{H}^{(-)}_{\rm int}(t) + \hat{H}^{(+)}_{\rm int}(t) ,
\end{eqnarray}
where $ C_{\rm int} $ is an interaction constant and $ \phi(k_S,k_I,t) $
describes a detailed space-time structure of the emitted photon
fields. The symbol $ {\rm H.c.} $ denotes Hermitian conjugate.
The operator $ \hat{H}^{(-)}_{\rm int} $
($ \hat{H}^{(+)}_{\rm int} $) stands for
the part of the interaction Hamiltonian $ \hat{H}_{\rm int} $
containing creation (annihilation) operators of modes in the
signal and idler fields.

The state of the signal and idler fields in the output
plane of the crystal is obtained solving the
Schr\"{o}dinger equation in the form:
\begin{eqnarray}      
 |\psi \rangle &=& \sum_{n=0}^{\infty} |\psi_n \rangle ,
   \nonumber \\
  |\psi_0 \rangle &=& |{\rm vac}\rangle , \nonumber \\
  |\psi_n \rangle &=& \left( - \frac{i}{\hbar} \right)^n
    \int_{-\infty}^{\infty} d\tau_1 \int_{-\infty}^{\tau_1}
    d\tau_2 \ldots \int_{-\infty}^{\tau_{n-1}} d\tau_n
    \nonumber \\
    & & \mbox{} \times
   \hat{H}_{\rm int}(\tau_1) \ldots \hat{H}_{\rm int}(\tau_n)
    |{\rm vac}\rangle , \hspace{1cm} n=1,2,\ldots .
   \nonumber \\
 & &
\end{eqnarray}
The signal and idler fields are assumed to be
in the vacuum state $ |{\rm vac}\rangle $ in the input
plane of the crystal.

If the number of photons in the signal and idler
fields is much lower than the number of independent modes constituting
these fields, we may approximately write:
\begin{eqnarray}      
   |\psi_n \rangle &\simeq & \left( - \frac{i}{\hbar} \right)^n
    \int_{-\infty}^{\infty} d\tau_1 \int_{-\infty}^{\tau_1}
    d\tau_2 \ldots \int_{-\infty}^{\tau_{n-1}} d\tau_n
   \nonumber \\
    & & \hspace{3mm}
    \hat{H}^{(-)}_{\rm int}(\tau_1) \ldots \hat{H}^{(-)}_{\rm int}(\tau_n)
    |{\rm vac}\rangle \nonumber \\
    &=& \left( - \frac{i}{\hbar} \right)^n \frac{1}{n!}
    \left[ \int_{-\infty}^{\infty} d\tau
    \hat{H}^{(-)}_{\rm int}(\tau) \right]^n |{\rm vac}\rangle
     , \nonumber \\
    & & \hspace{1cm} n=1,2,\ldots .
    \label{A3}
\end{eqnarray}
The state $ |\psi_n\rangle $ then describes the field containing
just $ n $ photon pairs.

Statistics of the generated photon pairs may be inferred from
the normally ordered moments of the
``creation operator of photon pairs''
$ \hat{P}_{\rm pair} $ and its Hermitian conjugate:
\begin{equation}        
 \hat{P}_{\rm pair}(\tau_S,\tau_I) =
 \underline{\hat{E}^{(-)}_S(\tau_{S}) \hat{E}^{(-)}_I(\tau_{I})} .
 \label{A4}
\end{equation}
The symbol
$ \hat{E}^{(+)}_l $ ($ \hat{E}^{(-)}_l $)
denotes the positive- (negative-) frequency part of
the electric-field amplitude of the signal ($ l=S $) and idler
($ l=I $) field:
\begin{eqnarray}           
 \hat{E}^{(+)}_l(\tau) &=& \sum_{k_l} e_l(k_l) \hat{a}_l(k_l)
 \exp (-i\omega_{k_l}\tau) ,
 \label{A5} \\
 \hat{E}^{(-)}_l &=& \left( \hat{E}^{(+)}_l \right)^\dagger
 \nonumber ;
\end{eqnarray}
$ e_l(k_l) $ is a normalization amplitude
of the mode $ k_l $, $ l=S,I $.
Underlining of the operators on the right-hand side of
Eq.~(\ref{A4})
means that ``only the signal and idler photons created in the
same elementary event are considered'' (see the expression
for $ |\psi_n\rangle $ in Eq.~(\ref{A3})).

We may write for the normally ordered moments of
$ \hat{P}^{\dagger}_{\rm pair} $ and $ \hat{P}_{\rm pair} $
in the framework of the above defined approximation:
\begin{eqnarray}    
 & &  \langle \psi_n | \left[ \prod_{j=1}^{n}
  \hat{P}_{\rm pair}^\dagger(\tau_{S_j},\tau_{I_j}) \right]
  \nonumber \\
 & & \hspace{1cm} \times
  \left[ \prod_{j=1}^{n} \hat{P}_{\rm pair}(\tau_{S_{n+1-j}},
  \tau_{I_{n+1-j}}) \right] |\psi_n\rangle = \nonumber \\
  & &  \left[ \prod_{j=1}^{n} \langle \psi_1 |
  \hat{P}_{\rm pair}^\dagger(\tau_{S_j},\tau_{I_j})
  |{\rm vac}\rangle \right]  \nonumber \\
  & & \hspace{1cm} \times \left[ \prod_{j=1}^{n} \langle {\rm vac}|
  \hat{P}_{\rm pair}(\tau_{S_{n+1-j}},\tau_{I_{n+1-j}})
  |\psi_1 \rangle \right] = \nonumber \\
  & & \prod_{j=1}^{n} \left| \langle \psi_1 |
  \hat{P}_{\rm pair}^\dagger(\tau_{S_j},\tau_{I_j})
  |{\rm vac}\rangle \right|^2 , \nonumber \\
   & & \hspace{3cm}    n=1,2,\ldots . \label{A6}
\end{eqnarray}
Assuming $ \tau_{S_1} = \tau_{S_2} = \ldots = \tau_S $ and
$ \tau_{I_1} = \tau_{I_2} = \ldots = \tau_I $, Eq.~(\ref{A6})
provides a simplified expression:
\begin{eqnarray}       
  & & \langle \psi_n | \left[ \hat{P}_{\rm pair}^\dagger(\tau_{S},\tau_{I})
   \right]^n \left[ \hat{P}_{\rm pair}(\tau_{S},
  \tau_{I}) \right]^n |\psi_n\rangle =
   \nonumber \\
  & & \hspace{2cm}
  \left| \langle \psi_1 |
  \hat{P}_{\rm pair}^\dagger(\tau_{S},\tau_{I})
  |{\rm vac}\rangle \right|^{2n} .
  \label{A7}
\end{eqnarray}
Assuming that the contribution from the state $ |\psi_n\rangle $
is much greater that those from the states $ |\psi_k\rangle $
for $ k=n+1,n+2, \ldots $, the relation in Eq.~(\ref{A7}) implies
that statistics of generated photon pairs is Poissonian.

If photon pairs are generated into one well-defined space-time mode,
than the positive-frequency part of the electric-field amplitude
$ \hat{E}^{(+)}_l $ of field $ l $ can be written as
\begin{equation}       
 \hat{E}^{(+)}_l(\tau) = a_l(\tau) \hat{a}_l ,
\end{equation}
where $ a_l $ describes amplitude per photon and
$ \hat{a}_l $ stands for annihilation operator of a photon
in field $ l $ ($ l=S,I $).
We then have:
\begin{eqnarray}       
 & & \langle \psi_n | \left[ \hat{P}_{\rm pair}^\dagger(\tau_{S},\tau_{I})
   \right]^n \left[ \hat{P}_{\rm pair}(\tau_{S},
  \tau_{I}) \right]^n |\psi_n\rangle = \nonumber \\
  & & | a(\tau_S) a(\tau_I) |^{2n}  \langle \psi_n |
   \left[ \underline{\hat{a}_S^{\dagger}\hat{a}_I^{\dagger}}
   \right]^n \left[ \underline{\hat{a}_S\hat{a}_I } \right]^n
   |\psi_n\rangle = \nonumber \\
  & & | a(\tau_S) a(\tau_I) |^{2n} n! = n!
   \left| \langle \psi_1 |
  \hat{P}_{\rm pair}^\dagger(\tau_{S},\tau_{I})
  |{\rm vac}\rangle \right|^{2n} .
  \label{A9}
\end{eqnarray}
Relations in Eq.~(\ref{A9}) are in agreement with Gaussian statistics
of generated photon pairs.

\end{document}